\RequirePackage{lineno} 
\documentclass[article,twocolumn]{revtex4}
\usepackage{graphicx}% Include figure files
\usepackage{dcolumn}% Align table columns on decimal point
\usepackage{bm}% bold math
\usepackage{color}
\usepackage{eso-pic}
\usepackage{amsmath}

\usepackage{xspace}
\newcommand{\LTO}{$\mathrm{LaTiO}_3$\xspace}
\newcommand{\STO}{$\mathrm{SrTiO}_3$\xspace}
\newcommand{\LAO}{$\mathrm{LaAlO}_3$\xspace}
\newcommand{\TiO}{$\mathrm{TiO}_2$\xspace}

\begin{document}

\title{\large Competition between electron pairing and phase coherence in superconducting interfaces}

\author{G. Singh$^{1,2 \dagger}$, A. Jouan$^{1,2 \dagger}$, L. Benfatto$^{3,4}$, F. Couedo$^{1,2}$, P. Kumar$^{5}$, A. Dogra$^{5}$, R. Budhani$^{6}$, S. Caprara$^{4,3}$, M. Grilli$^{4,3}$, E. Lesne$^{7}$, A. Barth\'el\'emy$^{7}$, M. Bibes$^{6}$, C. Feuillet-Palma$^{1,2}$, J. Lesueur$^{1,2}$, N. Bergeal$^{1,2*}$}

\affiliation{$^1$Laboratoire de Physique et d'Etude des Mat\'eriaux, ESPCI Paris, PSL Research University, CNRS, 10 Rue Vauquelin - 75005 Paris, France.\\
$^2$Universit\'e Pierre and Marie Curie, Sorbonne-Universit\'es,75005 Paris, France.\\
$^3$Institute for Complex Systems (ISC-CNR),  UOS Sapienza, Piazzale A. Moro 5, 00185 Roma, Italy\\
$^4$Dipartimento di Fisica Universit\`{a} di Roma``La Sapienza'', Piazzale A. Moro 5, I-00185 Roma, Italy.\\
$^5$National Physical Laboratory, Council of Scientific and Industrial Research (CSIR)Dr. K.S. Krishnan Marg, New Delhi-110012, India.\\
$^6$Condensed Matter Low Dimensional Systems Laboratory, Department of Physics, Indian Institute of Technology, Kanpur 208016, India.\\
$^7$Unit\'e Mixte de Physique CNRS-Thales, 1 Av. A. Fresnel, 91767 Palaiseau, France. }

\maketitle

\noindent $^\dagger$ : Both authors contributed equally to this work.\\
 $^*$ : Correspondence and request should be sent to N. B. (nicolas.bergeal@espci.fr).\\

%%%%%%%%%%%%
The large diversity of exotic electronic phases displayed by two-dimensional superconductors confronts physicists with new challenges. These include the recently discovered quantum Griffith singularity in atomic Ga films  \cite{xing}, topological phases in proximized topological insulators \cite{xu} and unconventional Ising pairing in transition metal dichalcogenide layers \cite{xi}. In  \LAO/\STO heterostructures, a gate tunable superconducting electron gas is confined in a quantum well at the interface between two insulating oxides \cite{Caviglia:2008p116}. Remarkably, the gas coexists with both magnetism \cite{li:2011p762,bert:2011p767} and strong Rashba spin-orbit coupling \cite{caviglia2,benshalom} and is a candidate system for the creation of Majorana fermions \cite{mohanta}. However, both the origin of superconductivity and the nature of the transition to the normal state over the whole doping range remain elusive. Missing such crucial information impedes harnessing this outstanding system for future superconducting electronics and topological quantum computing. Here we show that the superconducting phase diagram of \LAO/\STO is controlled by the competition between electron pairing and phase coherence. Through resonant microwave experiments, we measure the superfluid stiffness and infer the gap energy as a function of carrier density. Whereas a good agreement with the Bardeen-Cooper-Schrieffer (BCS) theory is observed at high carrier doping, we find that the suppression of $T_c$ at low doping is controlled by the loss of macroscopic phase coherence instead of electron pairing as in standard BCS theory. We find that only a very small fraction of the electrons condenses into the superconducting state and propose that this corresponds to the weak filling of a high-energy $d_{xz/yz}$ band, more apt to host superconductivity.\\

 \indent The superconducting phase diagram of  \LAO/\STO interfaces defined by plotting the critical temperature $T_c$ as a function of electrostatic doping has the shape of a dome. It ends into a quantum critical point, where the $T_c$ is reduced to zero, as carriers are removed from the interfacial quantum well \cite{Caviglia:2008p116, biscaras2}. Despite a few proposals \cite{maniv,gariglio}, the origin of this gate dependence and in particular the non-monotonic suppression of $T_c$ remains unclear.  There are two fundamental energy scales associated with superconductivity. On the one hand, the gap energy $\Delta$ measures the pairing strength between electrons that form Cooper pairs.  On the other hand, the superfluid stiffness $J_s$ determines the cost of a phase twist in the superconducting condensate. In ordinary BCS superconductors, $J_s$ is much higher than $\Delta$ and the superconducting transition is controlled by the breaking of Cooper pairs. However, when the stiffness is strongly reduced, phase fluctuations play a major role and the suppression of $T_c$  is expected to be dominated by the loss of phase coherence \cite{emery}.   Tunneling experiments in the low doping regime of \LAO/\STO interfaces evidenced the presence of a pseudogap in the density of states above $T_c$ \cite{richter}. This can be interpreted as the signature of pairing surviving above $T_c$ while superconducting coherence is destroyed by strong phase fluctuations, enhanced by a low superfluid stiffness \cite{bert2}. Superconductor-to-Insulator quantum phase transitions driven by gate voltage \cite{Caviglia:2008p116} or magnetic field \cite{biscaras4} also highlighted the predominant role of phase fluctuations in the suppression of $T_c$.  \\
  \indent  The low superfluid stiffness corresponds to a low superfluid density $n_s^\mathrm{2D}=\frac{4m}{\hbar^2}J_s$ which has to be analyzed within the context of the peculiar \LAO/\STO band structure. Under strong quantum confinement, the degeneracy of the $t_{2g}$ bands of \STO ($d_{xy}$, $d_{xz}$ and $d_{yz}$ orbitals) is lifted, generating a rich and complex band structure \cite{berner}.  Experiments performed on interfaces with different crystallographic orientations ([110] vs conventional [001] orientation) revealed the crucial role of orbitals hierarchy in the quantum well, and also suggested that only some specific bands could host superconductivity \cite{gervasi,joshua}. Here, we use a resonant microwave experiment to measure the kinetic inductance $L_k$ of the superconducting  \LAO/\STO interface. This allows us to determine the evolution of the superfluid stiffness $J_s=\frac{\hbar^2}{4e^2L_k}$ and corresponding superfluid density $n_s^\mathrm{2D}$ in the phase diagram. \\
            \begin{figure}[t]
\includegraphics[width=8.5cm]{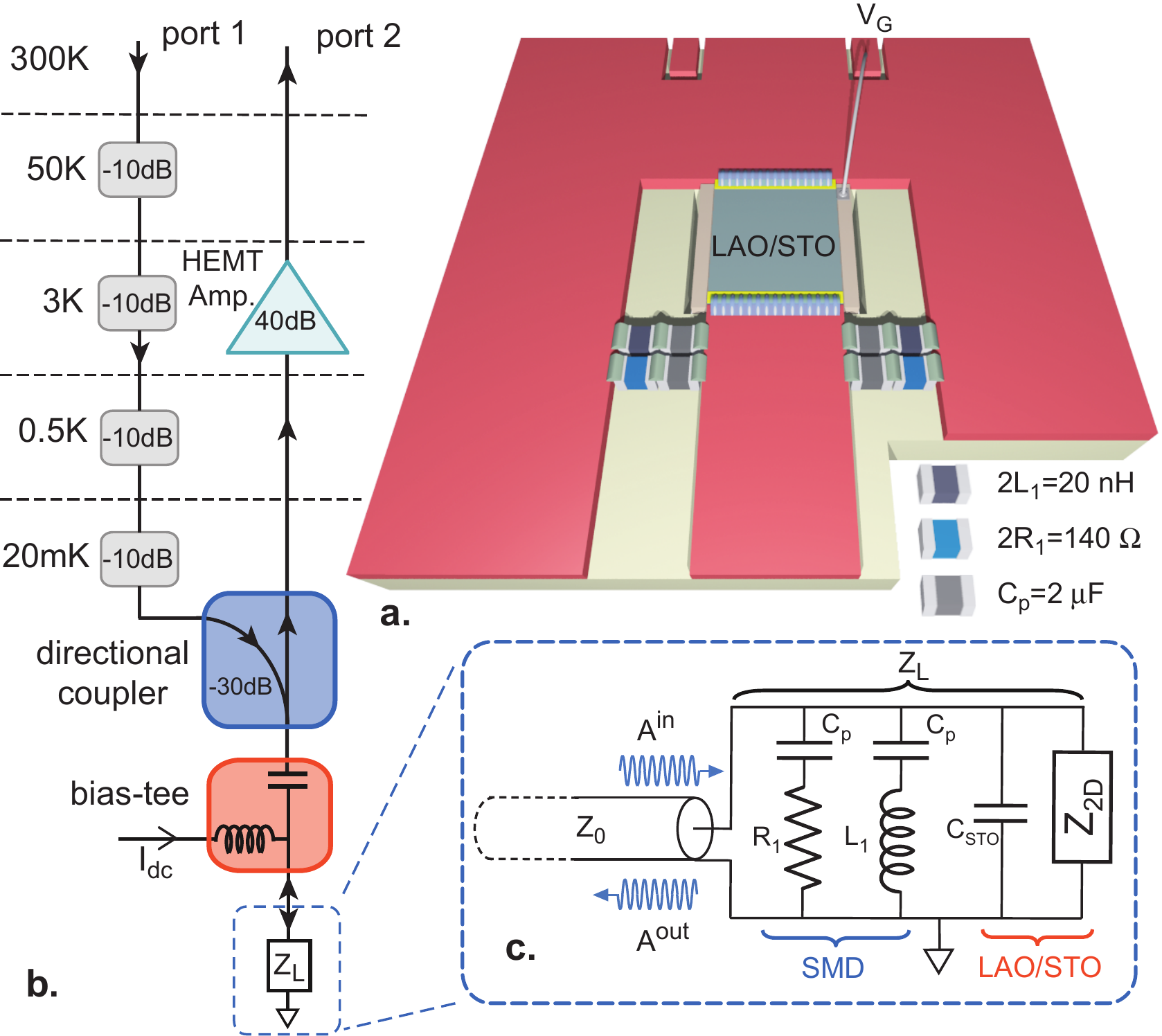}
\caption{The \LAO/\STO sample and its microwave measurement set-up. a) \LAO/\STO sample inserted between the central strip and the ground of a CPW transmission line, in parallel with SMD inductors $L_1$ and resistors $R_1$. $C_p$ are protective capacitors that avoid dc current to flow through $L_1$ and $R_1$ without influencing $\omega_0$. b) Sample circuit of impedance $Z_L$ in its microwave measurement that includes an attenuated input line and an amplified output line separated by a directional coupler. A bias-tee allows dc biasing of the sample. c) Equivalent electrical circuit of the sample circuit including the SMDs and the \LAO/\STO hetero-structure modeled by a capacitor $C_\mathrm{STO}$ and an impedance $Z_\mathrm{2D}$. }
\end{figure}

%%%%%%

 \begin{figure}[b]
\includegraphics[width=9cm]{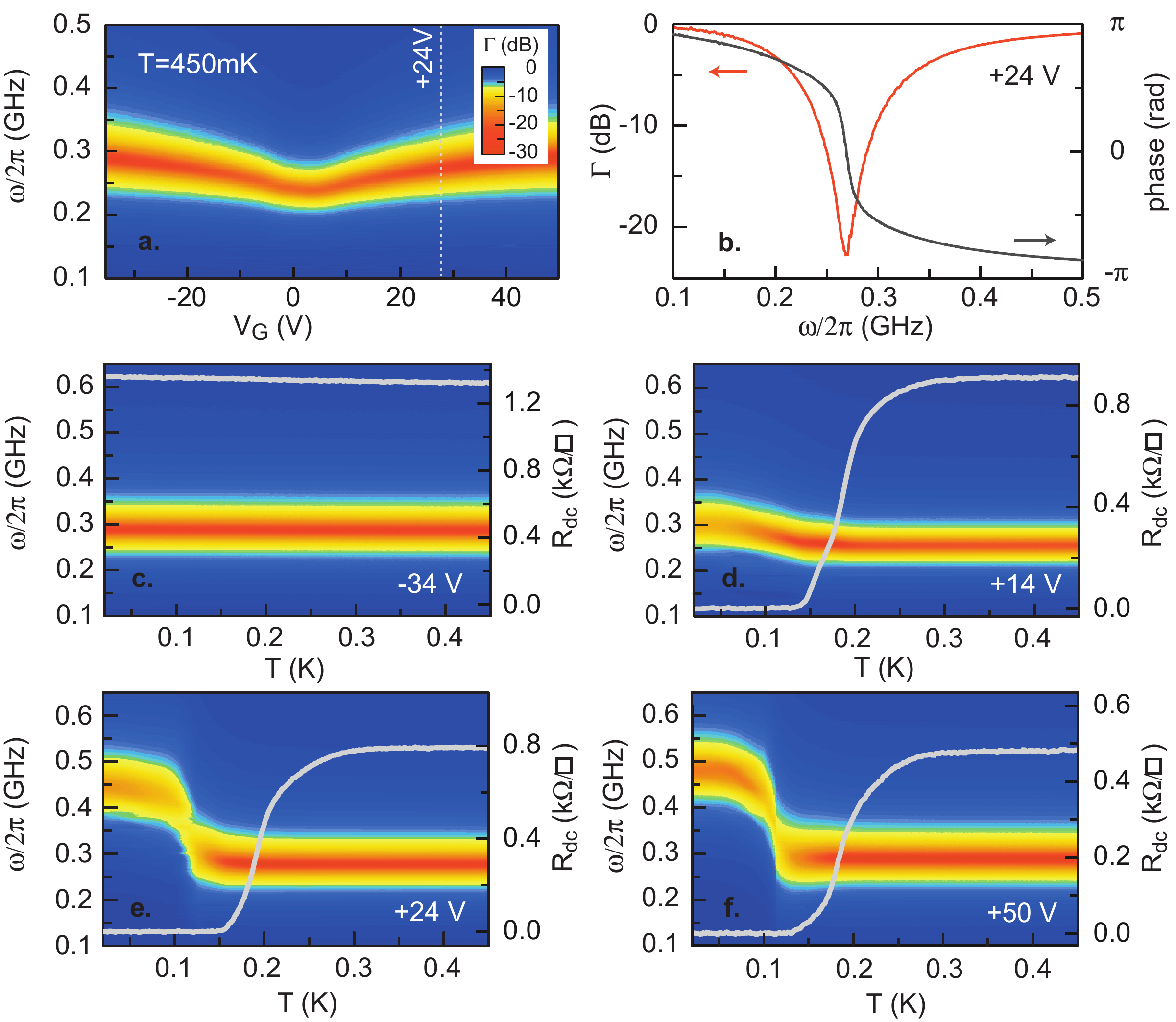}
\caption{Resonance shift in the superconducting state. a) $\Gamma(\omega)$ in dB (color scale) as a function of frequency and $V_\mathrm{G}$ at T=450mK. b) Amplitude (left axis) and phase (right axis) of $\Gamma(\omega)$ showing the resonance frequency for $V_\mathrm{G}$= +24V at T=450mK. c,d,e,f) $\Gamma(\omega)$ in dB (color scale) as a function of frequency and temperature for the selected gate values, $V_\mathrm{G}$=-34V (a), $V_\mathrm{G}$=+14V (b), $V_\mathrm{G}$=+24V (c), $V_\mathrm{G}$=+50V (d). The corresponding dc resistance as a function of temperature is shown on the right axis.}
\end{figure}

Figure 1 gives a schematic description of our experimental set-up, largely inspired by recent developments in the field of quantum circuits \cite{wallraff,bergeal}.  The \LAO/\STO sample is mounted on a microwave circuit board which is anchored to the 18 mK cold stage of a dilution refrigerator. It is embedded into a RLC resonant circuit whose inductor L$_1$ and resistor R$_1$ are Surface Mounted microwave Devices, and whose capacitor $C_\mathrm{STO}$ is the \STO substrate in parallel with the two-dimensional electron gas (2-DEG)  (Fig. 1a and 1c.). After calibration, the measurement of the complex reflection coefficient  $\Gamma(\omega)=\frac{A^\mathrm{out}}{A^\mathrm{in}}$ at the input of the resonant sample circuit allows to determine the complex conductance  $G(\omega)=G_1(\omega)-iG_2(\omega)$ of the  2-DEG in a frequency band centered on the resonance frequency $\omega_0=\frac{1}{\sqrt{L_1C_\mathrm{STO}}}$ (see Methods). In the normal state ($T>T_c$),  C$_\mathrm{STO}$ is deduced from $\omega_0$  for each gate value (Fig. 2a,b). In the superconducting state, the 2-DEG conductance acquires an imaginary part $G_2(\omega)=\frac{1}{L_k\omega}$ that modifies $\omega_0$, as the total inductance is then given by $L_1$ in parallel with $L_k$.  The superconducting transition observed in dc  resistance ($R_{dc}$=0 $\Omega$) for positive gate voltages $V_\mathrm{G}$, coincides with a continuous shift of $\omega_0$ towards high-frequency (Fig. 2d,e,f). In absence of superconductivity (for $V_\mathrm{G}$ $<$ 0 V), the resonance frequency remains unchanged (Fig 2c). \\

 We now determine the gate dependence of the important energy scales in superconducting \LAO/\STO interfaces, and compare them with the BCS theory predictions.  
 In Figure  3a, we show the experimental superfluid stiffness  $J_s^\mathrm{exp}=\frac{\hbar^2}{4e^2L_k}$  as a function of $V_\mathrm{G}$ at the lowest temperature $T$ = 20 mK ($\simeq$ 0 K in the following). For a single band BCS superconductor, within a dirty limit approximation ($l$-mean free path- $<$ $\xi$ -coherence length-) and for $\omega\ll\Delta/\hbar$, $J_s$ can be expressed as a function of the gap energy \cite{pracht} : 
\begin{eqnarray}
J_s(T\simeq0)=\frac{\pi\hbar}{4e^2k_BR_n}\cdot \Delta(T\simeq0)
\end{eqnarray}
where $R_n$ = $R(T\gtrsim T_c)$ is the normal state resistance (inset Fig. 3b). A remarkable agreement is obtained between experimental data ($J_s^\mathrm{exp}$) and BCS prediction ($J_\mathrm{BCS}$) in the overdoped (OD) regime defined by $V_\mathrm{G}$ $>$ $V_\mathrm{G}^\mathrm{opt}$ $\simeq$ 27 V, assuming in Eq. (1) a  gap energy  $\Delta=\Delta_\mathrm{BCS}$ = 1.76$k_BT_c$ (Fig. 3a). In this regime, the superfluid stiffness $J_s^\mathrm{exp}$ takes a value much higher than $T_c$ in agreement with the BCS paradigm.  However, in the underdoped (UD) regime, corresponding to $V_\mathrm{G}$ $<$ $V_\mathrm{G}^\mathrm{opt}$, a discrepancy between the data and the BCS calculation is observed.  The superfluid stiffness $J_s^\mathrm{exp}$ drops significantly while $T_c$ and $J_\mathrm{BCS}$ evolve smoothly before vanishing only when approaching closely the quantum critical point where $T_c$  $\simeq$ 0 ($V_\mathrm{G}$= 4 V). This indicates that the global phase coherence of the superconducting condensate is partially lost in the 2-DEG. Such behavior is due to strong phase fluctuations, probably reinforced by the presence of spatial inhomogeneities which has been proposed as an explanation for the observed broadening of the superconducting transitions \cite{sergio}. In this context, it was shown that the 2-DEG in \LAO/\STO interfaces exhibits the physics of a Josephson junction array consisting of superconducting islands coupled through a metallic 2-DEG \cite{biscaras4,guenevere}. Whereas in the OD regime the islands are robust and well connected (homogeneous-like),  in the UD regime, the charge carrier depletion makes the 2-DEG more inhomogeneous. In this case, the system can maintain a rather high $T_c$ ($R_\mathrm{dc}$ = 0 $\Omega$) as long as the dc current can percolate between islands. However, as a fraction of the interface is non-superconducting, the overall stiffness $J_s^\mathrm{exp}$ is lower than the one expected in a homogenous system of similar $T_c$.\\

         \begin{figure}[t]
\includegraphics[width=8.5cm]{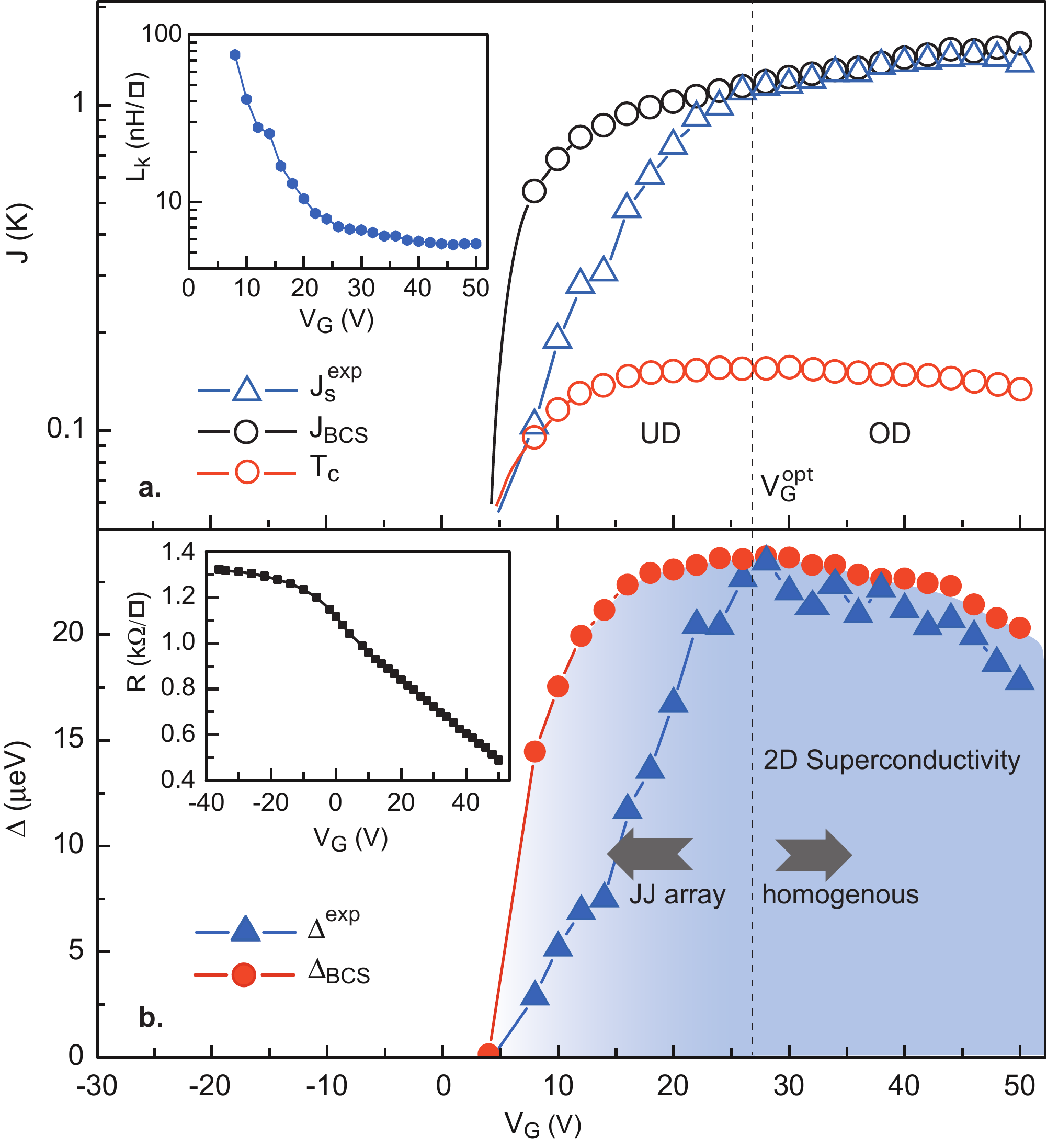}
\caption{Superfluid stiffness and phase diagram. a) Experimental superfluid stiffness $J_s^\mathrm{exp}$  as a function of $V_\mathrm{G}$ compared with $T_c$ (taken at R=0 $\Omega$) and with the BCS theoretical stiffness $J_\mathrm{BCS}$. Inset) $L_k$ as a function of $V_\mathrm{G}$. b) Superfluid stiffness converted into a gap energy $\Delta_s^\mathrm{exp}$ as a function of $V_\mathrm{G}$ compared with the BCS gap energy $\Delta_\mathrm{BCS}$. Inset) Normal sheet resistance as a function of $V_\mathrm{G}$.}
\end{figure}

Using Eq. (1), we now convert $J_s^\mathrm{exp}$ into a gap energy $\Delta^\mathrm{exp}_s$ and compare it directly with $\Delta_\mathrm{BCS}$ = 1.76$k_BT_c$ (Fig. 3b). Strikingly, these two characteristic energy scales of superconductivity evolve with doping quite differently. While $J_s^\mathrm{exp}$ continuously increases with $V_\mathrm{G}$, $\Delta^\mathrm{exp}_s$ has a dome shape dependence. More precisely, in the OD regime, $\Delta^\mathrm{exp}_s$ coincides with the BCS value and decreases like $T_c$ while the superfluid stiffness increases : this is a clear indication that $T_c$ is controlled by the pairing energy ($\propto J_s^\mathrm{exp} R_n$) as in the BCS scenario. On the contrary, in the UD part of the phase diagram, $\Delta^\mathrm{exp}_s$ departs from $\Delta_\mathrm{BCS}$. The maximum energy gap at optimal doping ($V_\mathrm{G}^\mathrm{opt}$ $\simeq$  27 V) is $\Delta^\mathrm{exp}_s$ $\approx$ 23 $\mu eV$. By using tunneling spectroscopy on planar Au/\LAO/\STO junctions, \textit{Richter et al.} have reported an energy gap in the density of states of $\sim40$ $\mu$eV  for optimally doped \LAO/\STO interfaces \cite{richter}, which corresponds to $\Delta_{BCS}\simeq1.7k_BT_c$ in agreement with our result.  However, the tunneling gap was found to increase in the UD regime, which is different from the behavior of $\Delta^{\mathrm{exp}}_s$ reported here.  In addition, a pseudogap has been observed above $T_c$ in this regime, as also reported in High-$T_c$ superconducting cuprates \cite{timusk} or in strongly disordered films of conventional superconductors \cite{sacepe,pracht}. The results obtained by the two experimental approaches can be reconciled by considering carefully the measured quantities. In our case, the superconducting gap $\Delta^\mathrm{exp}_s$ probed by microwaves is directly converted from the stiffness of the superconducting condensate and is therefore only reflective of the presence of a true phase-coherent state. On the other hand, tunneling experiments probe the single particle density of states and can evidence pairing even without phase coherence. The two experimental methods provide complementary informations which indicate that in the UD region of the phase diagram, the superconducting transition is dominated by the loss of phase coherence rather than the pairing. In the region $V_\mathrm{G}<0$, some non-connected superconducting islands could already exist without contributing to the macroscopic stiffness of the 2-DEG.\\
         \begin{figure}[t]
\includegraphics[width=8.5cm]{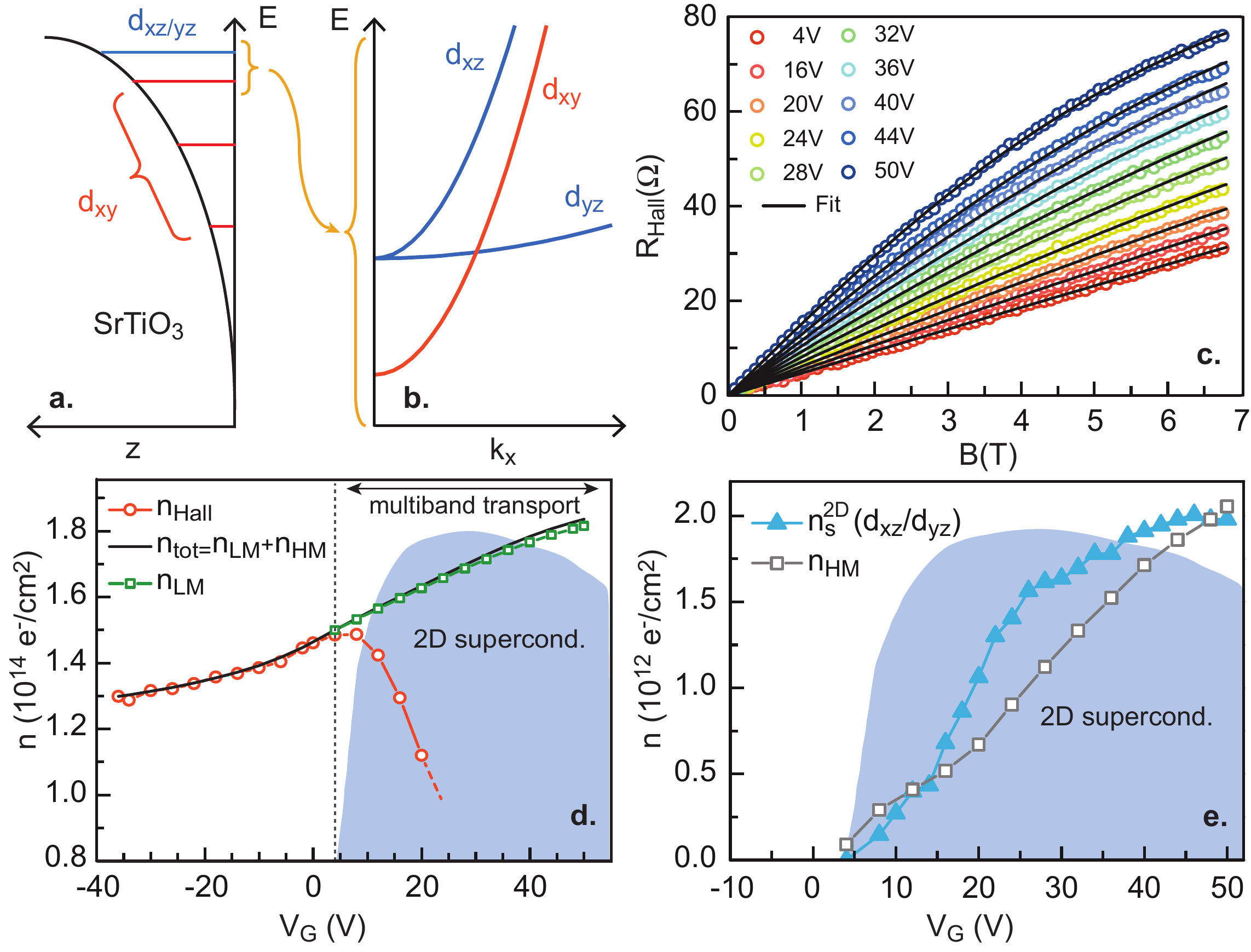}
\caption{Superfluid density and Hall effect analysis. a) Scheme of the interfacial quantum well showing the splitting of the $t2g$ bands. b) Simplified scheme of the band structure taking into account only the last filled $d_{xy}$ subband, the $d_{xz}$ band and the $d_{yz}$  band. c) Hall resistance as a function of magnetic field for different gate voltages fitted by at two-band model (see Supplementary Information). d) Hall carrier density $n_\mathrm{Hall}=\frac{B}{eR_\mathrm{Hall}}$ extracted in the limit $B\rightarrow$ 0 and LMC density $n_\mathrm{LM}$ extracted from the two-band analysis.  The total carrier density $n_\mathrm{tot}$  is obtained by matching the charging curves of the gate capacitance with $n_\mathrm{Hall}$ at negative $V_\mathrm{G}$. The unscaled  $T_c$ dome in the background indicates the region where superconductivity is observed. e)  Superfluid density $n_s^\mathrm{2D}$ calculated from $J_s^\mathrm{exp}$ using a mass $m_{xz/yz}$, compared with the HMC density $n_\mathrm{HM}$.}
\end{figure}
\indent  A simplified scheme of the band structure in the interfacial quantum well is presented in Figures 4a and 4b. The degeneracy of the three $t_{2g}$ bands is lifted by  confinement in the $z$ direction, leading to a splitting that is inversely proportional to the effective masses $m_z$ along this direction.
$d_{xy}$ subbands are isotropic in the interface plane with an effective mass $m_{xy}$=0.7$m_0$ whereas the $d_{xz}$/$d_{yz}$ bands are anisotropic with a corresponding average mass $m_{xz/yz}=\sqrt{m_xm_y}\simeq$ 3.13$m_0$. At low carrier densities, we expect several $d_{xy}$ subbands to be populated, whereas at higher density ($V_\mathrm{G}$ $>$ 0 V), the Fermi energy should enter into the $d_{xz}$/$d_{yz}$ bands, leading to multiband transport. Recent measurements of quantum oscillations showed that, in addition to a majority of low-mobility carriers (LMC), a small amount of high-mobility carriers (HMC) is also present, with an effective mass close to the $m_{xz/yz}$ one \cite{yang}. Despite a band mass substantially higher than the $m_{xy}$ one, these carriers acquire a high-mobility as $d_{xz/yz}$ orbitals extend deeper in \STO where they recover bulk-like properties, including reduced scattering, higher dielectric constant and better screening. Multiband transport was also evidenced in Hall effect measurements \cite{biscaras2,Kim:2010p9791}. Whereas the Hall voltage is linear in magnetic field $B$ in the UD regime corresponding to one-band transport, this is not the case in the OD regime because of the contribution of a new type of carriers (the HMC). We performed a two-band analysis of the Hall effect data combined with gate capacitance measurement to determine the contribution of the two populations of carriers to the total density $n_\mathrm{tot}$ (Fig. 4c) \cite{biscaras2}. The first clear signature of multiband transport is seen when the Hall carrier density $n_\mathrm{Hall}$, measured in the limit B$\rightarrow$0, starts to decrease with $V_\mathrm{G}$ instead of following the charging curve of the capacitance ($n_\mathrm{tot}$ in Fig. 4d). Figures 4d and 4e show that LMC of density $n_\mathrm{LM}$ are always present, whereas a few HMC of density $n_\mathrm{HM}$ are injected in the 2-DEG for positive $V_\mathrm{G}$, which corresponds to the region of the phase diagram where superconductivity is observed. In consistency with quantum oscillations measurements, we identify the LMC and the HMC as coming from the $d_{xy}$ and $d_{xz/yz}$ subbands respectively and we emphasize that the addition of HMC in the quantum well triggers superconductivity.\\

To further outline the relation between HMC and superconductivity, we extract the superfluid density $n_s^\mathrm{2D}$ from $J_s^\mathrm{exp}$ assuming a mass $m_{xz/yz}$ for the electrons, and plot it as a function of the gate voltage (Fig. 4e). It increases continuously to reach $n_s^\mathrm{2D}\simeq~2~\times~10^{12}~e^{-}\cdot\mbox{cm}^{-2}$ at maximum doping, which is approximately $1\%$ of the total carrier density. This behavior is similar to the one observed for the superfluid density measured by scanning SQUID experiments \cite{bert2}. The comparison of $n_s^\mathrm{2D}$ with $n_\mathrm{HM}$ in Fig. 4e shows that, unexpectedly, both quantities have a very similar dependence with the gate voltage and almost coincide numerically. This suggests that the emergence of the superconducting phase is related to the filling of $d_{xz/yz}$ bands, whose high density of states is favorable to superconductivity. This is consistent with the observation of a gate-independent superconductivity in [110] oriented \LAO/\STO interfaces for which the $d_{xz/yz}$ bands have a lower energy than the $d_{xy}$ subbands and are therefore always filled \cite{gervasi}. 
The fact that  $n_s^\mathrm{2D}\simeq n_\mathrm{HM}$ is somewhat intriguing as the dirty limit that we used in Eq. (1) implies that $n_s^\mathrm{2D}$ should correspond to a fraction of the total normal carrier density (approximately 2$\Delta\tau/\hbar$, where $\tau$ is the scattering time) and not to $n_\mathrm{HM}$.
To understand such apparent discrepancy, it is needed to go beyond single-band superconductor models that can not account correctly for the unusual $t_{2g}$-based interfacial band structure of \LAO/\STO interfaces.  Further investigations of recent experimental \cite{dai} and theoretical \cite{chubukov} developments on superconductors having two dissimilar bands (\textit{e. g.} clean and dirty, weak and strong coupling),  should provide an appropriate framework to address this question.\\

{\textbf{Methods}}\\

\textbf{Sample growth and gate deposition.}\\

\indent In this study, we used 8 uc thick \LAO epitaxial layers grown on 3$\times$3~mm$^2$ \TiO -terminated [001] \STO single crystals by Pulsed Laser Deposition. The
substrates were treated with buffered HF to expose TiO$_2$ terminated surface. Before deposition, the substrate was heated to 830 $^\circ$C for one hour in an oxygen
pressure of 7.4$\times$10$^{-2}$ mbar.  The thin film was deposited at 800 $^\circ$C in an oxygen partial pressure of 1$\times$10$^{-4}$mbar. The \LAO target was ablated with a KrF excimer laser at a rate of 1Hz with an energy density of 0.56-0.65 Jcm$^{-2}$. The film growth mode and thickness were monitored using RHEED (STAIB, 35 keV) during deposition.  After the growth, a weakly conducting metallic back-gate of resistance $\sim$100 k$\Omega$ (to avoid microwave short cut of the 2-DEG) is deposited on the backside of the 100 $\mu m$ thick \STO substrate.\\

\textbf{Complex conductivity and kinetic inductance of a superconductor.} \\

 \indent In superconducting thin films, $J_s$ is usually assessed either from penetration depth measurements \cite{ganguly} or from dynamic transport measurements \cite{kitano,scheffler}. This latter method was adapted in this work for the specific case of \LAO/\STO samples which requires the use of a low-temperature dilution refrigerator. While superconductors have an infinite dc conductivity, they exhibit a finite complex conductivity $\sigma(\omega)$ at non-zero frequency,  which in 2D translates into a sheet conductance  $G(\omega)=G_1(\omega)-iG_2(\omega)$. The real part $G_1(\omega)$  accounts for the transport of unpaired electrons existing at $T\neq0$ and  $\omega\neq0$,  and the imaginary part $G_2(\omega)$ accounts for the transport of Cooper pairs. The expression of $G_1(\omega)$  and $G_2(\omega)$  have been derived by Mattis and Bardeen in a seminal paper which gives a complete description of the electrodynamic response of superconductors based on the BCS theory \cite{MB}. In the limit $\hbar\omega\ll\Delta$, which is well satisfied here ($\Delta\simeq$ 5GHz), a superconductor behaves essentially as an inductor and $G_2(\omega)$=$\frac{1}{L_k\omega}$, where $L_k$ is the kinetic inductance of the superconductor due to the inertia of Cooper pairs \cite{tinkham}. In our experiment, $L_k$ corresponds to a sheet inductance (i.e for a square sample). Below $T_c$ the total inductance is given by $L_t(T)=\frac{L_1L_k(T)}{L_1+L_k(T)}$ corresponding to the kinetic term $L_k$ in parallel with the constant SMD inductance $L_1$. Notice that in our circuit, the geometric inductance of the sample is negligible compared to the kinetic one. As for $T<T_c$, $L_k$ decreases when lowering the temperature, the superconducting transition observed in dc  resistance for positive gate voltages, coincides with a continuous shift of $\omega_0$ towards high-frequency (Fig. 3).\\
 
\textbf{ Microwave reflection coefficient }\\

       \indent In a microwave circuit, the reflection coefficient at a discontinuity of a transmission line is defined as the ratio of the complex amplitude of the reflected wave $A^\mathrm{out}(\omega)$ to that of the incident wave $A^\mathrm{in}$. When the transmission line is terminated by a load of impedance  $Z_L(\omega)$, it is given by \cite{pozar}
\begin{eqnarray}
\Gamma(\omega)=\frac{A^\mathrm{out}(\omega)}{A^\mathrm{in}(\omega)}=\frac{Z_L(\omega)-Z_0}{Z_L(\omega)+Z_0} 
\label{reflect}
\end{eqnarray}
 where  $Z_0$ = 50$\Omega$ is the characteristic impedance of a standard microwave line. The measurement of $\Gamma(\omega)$ allows therefore to access directly to the load impedance $Z_L(\omega)$ or equivalently its admittance $G_L(\omega)=1/Z_L(\omega)$, commonly called complex conductance. In this work, a \LAO/\STO heterostructure is inserted between the central strip of a coplanar waveguide guide (CPW) transmission line and its ground, and is electrically connected through negligible contacts impedance. The high dielectric constant of the \STO substrate at low temperature (i. e. $\epsilon\simeq$ 24000) generates a sizable capacitance $C_\mathrm{STO}$ in parallel with the 2-DEG which has to be correctly subtracted to extract the dynamic transport properties of the 2-DEG. This problem can be overcome by embedding the \LAO/\STO heterostructure in a RLC resonating circuit whose inductor L$_1$=10nH and resistor R$_1$=70 $\Omega$ are Surface Mounted microwave Devices (SMD), and whose capacitor $C_\mathrm{STO}$ is the \STO substrate in parallel with the 2-DEG  (Figure 1a and 1c.).  A directional coupler allows to send the microwave signal from port 1 to the sample through a bias-tee, and to separate the reflected signal which is amplified by a low-noise cryogenic HEMT amplifier before reaching port 2 (Fig. 1b). Such type of microwave reflection set-up has been widely used in the quantum circuit community. \\
 \indent  After cooling the sample to 3K, the back-gate voltage is first swept to its maximum value +50V while keeping the 2-DEG at the electrical ground, to insure that no hysteresis will take place upon further gating \cite{biscaras3}. The transmission coefficient $S_{21}(\omega)$ between the two ports is measured with a vector network analyzer. The reflection coefficient $\Gamma(\omega)$ taken at the discontinuity between the CPW line and the circuit formed by the sample and SMD components (see Fig 1) is given by Eq. (\ref{reflect}) where $Z_\mathrm{L}=1/G_\mathrm{L}$ and $G_\mathrm{L}$ is obtained by summing up all the admittances in parallel in the RLC circuit of Fig 1c. Loses of \STO substrate are not included in this model as they only renormalize the amplitude of the absorption deep without modifying the resonance frequency (or equivalently $G_2(\omega)$). Standard microwave network analysis relates $\Gamma(\omega)$ to the measured $S_{21}(\omega)$, through complex error coefficients representing the reflection tracking, the source match and the directivity coefficient of the set-up \cite{pozar}.  A precise calibration procedure requiring three reference impedances, usually an open, a short and a match standard, allows a complete determination of these error coefficients. In this experiment, the microwave set-up was calibrated by using as references, the impedances of the sample circuit in the normal state of the 2-DEG for different gate values.\\

\textbf{References for Methods section}

%%%%%%%%%

\textbf{Acknowledgments}\\
We acknowledge  C. Castellani and J. Lorenzana for useful discussions. This work has been supported by the R\'egion Ile-de-France in the framework of CNano IdF, OXYMORE and Sesame programs, by CNRS through a PICS program (S2S) and ANR JCJC (Nano-SO2DEG). Part of this work has been supported by the IFCPAR French-Indian program (contract 4704-A). Research in India was funded by the CSIR and DST, Government of India.\\

\end{document}